\documentclass[fleqn]{elsart3-1}

 \usepackage{graphicx}
 \usepackage{psfrag}

\usepackage{amssymb}
\usepackage{amsmath}
\usepackage[figuresright]{rotating}
\newcommand{\bm}{\boldsymbol}
\newcommand{\nablam}{\bm{\nabla}}

\usepackage[english,francais]{babel}

\def\og{\leavevmode\raise.3ex\hbox{$\scriptscriptstyle\langle\!\langle$~}}
\def\fg{\leavevmode\raise.3ex\hbox{~$\!\scriptscriptstyle\,\rangle\!\rangle$}}

\begin{document}

\begin{frontmatter}


\selectlanguage{english}
\title{Nonlocal description of sound propagation through an array of Helmholtz resonators}


\selectlanguage{english}
\author[1]{Navid Nemati},
\ead{nnemati@mit.edu}
\author[1]{Anshuman Kumar},
\ead{akumr@mit.edu}
\author[2]{Denis Lafarge},
\ead{denis.lafarge@univ-lemans.fr}
\author[1]{Nicholas X. Fang}
\ead{nicfang@mit.edu}

\address[1]{Department of Mechanical Engineering, Massachusetts Institute of Technology, 77 Massachusetts Avenue,  Cambridge, MA 02139, USA}
\address[2]{Laboratoire d'Acoustique de l'Universit\'e du Maine, UMR 6613, Avenue Olivier Messiaen, 72085 Le Mans Cedex 9, France}


\medskip

\begin{abstract}
A generalized macroscopic nonlocal theory of sound propagation in 
rigid-framed porous media saturated with a viscothermal fluid has 
been recently proposed, which takes into account both temporal and spatial dispersion.  
Here, we consider applying this theory capable to describe resonance effects, to the case of sound propagation through an array of Helmholtz resonators whose unusual metamaterial properties such as negative bulk  modulii, have been experimentally demonstrated. Three different calculations are performed, validating the results of the nonlocal theory, relating to the frequency-dependent Bloch wavenumber and bulk modulus of the first normal mode, for 1D propagation in 2D or 3D periodic structures.

\vskip 0.5\baselineskip

\selectlanguage{francais}
\noindent{\bf R\'esum\'e}
\vskip 0.5\baselineskip
\noindent
{\bf Description nonlocale de la propagation du son dans une chaine de r\'{e}sonateurs de Helmholtz. }
Une th\'eorie macroscopique nonlocale g\'en\'erale de la propagation du son dans les milieux poreux \`{a} structure rigide satur\'es par un fluide viscothermique a r\'{e}cemment vu le jour. Tenant un compte complet des dispersions tant temporelles que spatiales, elle d\'ecrit enti\`{e}rement  les r\'esonances. Ici, nous l'appliquons au cas de la propagation du son dans un r\'eseau de r\'esonateurs de Helmholtz, dont les propri\'et\'es non usuelles (modules de compressibilit\'e n\'egatifs) ont \'et\'e  \'etablies exp\'erimentalement. Trois calculs diff\'erents sont pr\'esent\'es, qui valident les r\'esultats de la th\'eorie nonlocale, relatifs au nombre d'onde et module de compressibilit\'e fonctions de la fr\'equence, du mode de Bloch principal (le moins att\'enu\'e), pour une propagation 1D en g\'eom\'etries p\'eriodiques 2D ou 3D.

\vskip 0.5\baselineskip
\noindent{\small{\it Mots-cl\'es~:} Résonateur d'Helmholtz; Metamatériaux acoustiques; Description nonlocale; Dispersion spatiale; Fluide viscothermique; Module de compressibilité négatif}}
\end{abstract}
\end{frontmatter}

\selectlanguage{francais}

\selectlanguage{english}
\section{Introduction}
\label{introduction}

\begin{figure}[h!]
\centering

\includegraphics[width=7.5cm]{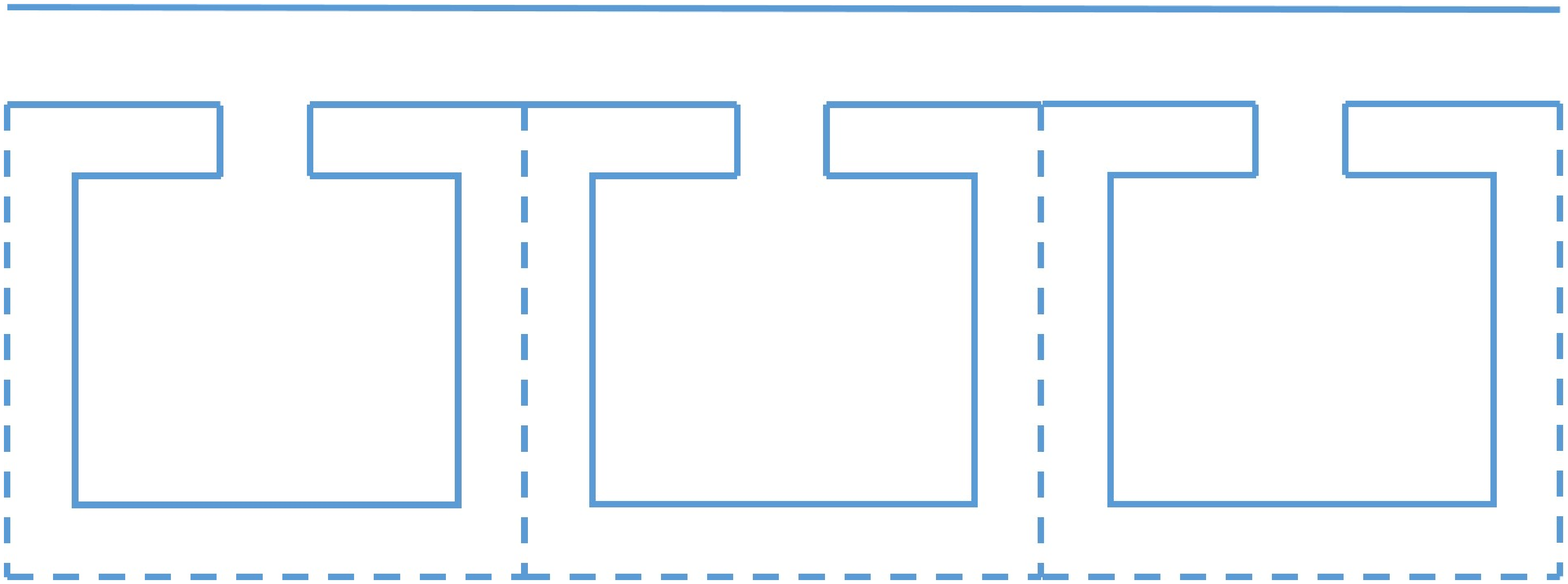} 
\hspace*{1cm}
\includegraphics[width=5.5cm]{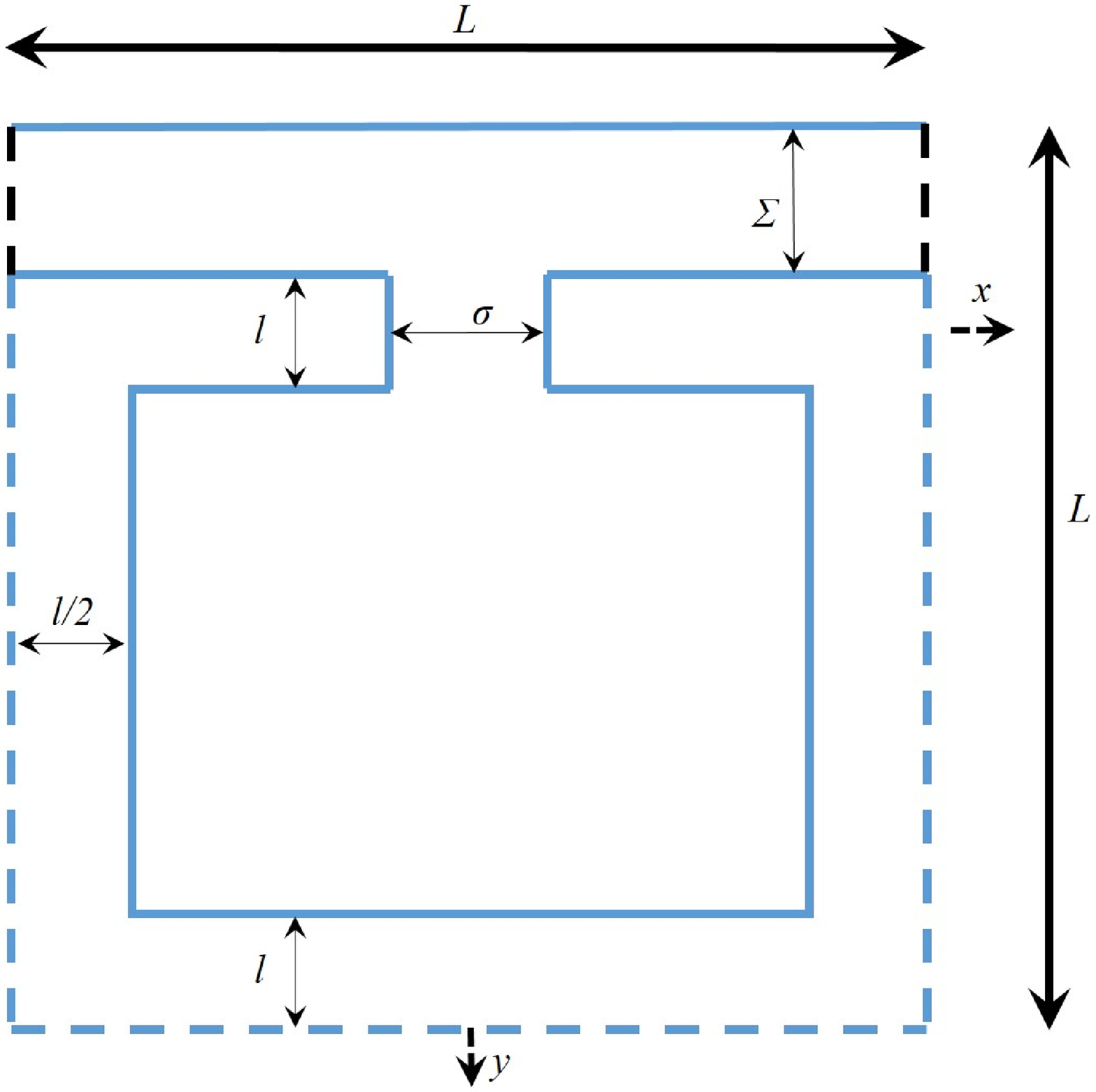} 
\caption{Left: illustration of a 2D array of Helmholtz resonators. Right: a periodic cell of the structure, with $L=1cm$, $\Sigma=0.2L$, $\sigma=0.015L$, and $l=0.15L$.}
\label{geometry}
\end{figure}

We employ here a generalized macroscopic nonlocal theory of sound propagation in rigid-framed porous media saturated with a viscothermal fluid \cite{lafarge2013} to describe the behavior of an acoustic metamaterial made of an array of Helmholtz resonators filled with air (see Fig. \ref{geometry} left). Inspired by the electromagnetic theory and a thermodynamic consideration relating to the concept of acoustic part of energy current density, this theory  allows to go beyond the limits of the classical local theory and within the limits of linear theory, to take into account not only temporal dispersion, but also spatial dispersion. In the framework of the new approach, an homogenization procedure is proposed, through solving two independent microscopic action-response problems each of which related to the effective density and effective bulk modulus of the material. Contrary to the classical (two-scale asymptotic) method of homogenization, there is no length-constraint to be considered alongside of the development of the new method, thus, there would be no frequency limit for the medium effective  properties to be valid, and also materials with different length scales can be treated. The homogenization procedure offers a systematic way of obtaining the effective properties of the materials, regardless of their geometries. These characters of the nonlocal approach give the possibility to describe the porous media with specific geometries causing  metamaterial behavior. A metamaterial with periodic structure will be studied: two-dimensional and three dimensional chain of Helmholtz resonators connected in series.  

By the local theory we refer to space locality. Nonlocality in time, or temporal dispersion, has been already taken into account through models for wave propagation in porous media \cite{zwikker,johnson,champoux,lafarge97}. That means, in Fourier space the effective density and bulk modulus depend on the frequency $\omega$. In other terms, the field dynamics at one location retains a memory of the field values at this location but is not affected by the neighboring values.  The local description is usually based on retaining only the leading order terms in the two-scale  homogenization method \cite{burridge,norris,zhou,smeulders,auriault80,auriault2009,lafarge97} using an asymptotic two-scale approach in terms of a characteristic length of the medium, the period $L$ in periodic media, which is supposed to be much smaller than the wavelength $\lambda$ \cite{sanchez,bensoussan}. Efforts have been performed to extend the asymptotic method of homogenization to  higher frequencies for the periodic composite materials \cite{kaplunov,craster1} and rigid porous media \cite{boutin2012-mod} by introducing another type of scale separation to which the asymptotic multi-scale procedure applies. Enhanced asymptotic method has been adapted to describe  sound propagation in rigid porous media with embedded damped Helmholtz resonators \cite{boutin2013} exhibiting scattering different  from Bragg scattering at high frequency in periodic media. 

An effective medium approach has been proposed for periodic elastic composites based on surface responses of a structural unit of the material \cite{sheng}, which can describe the macroscopic parameters beyond the frequencies within the long wavelength limit. Unlike the classical methods, based on the introduction of two-scale asymptotic expansions, or coherent potential approximation \cite{CPA} based on  the effective-medium parameters  minimizing scatterings in the long-wavelength limit, the homogenization scheme presented in \cite{sheng} uses matching the lowest-order scattering amplitudes arising from a periodic unit cell of the metamaterial with that of a homogenized material. As such, local resonant scattering can be captured as well by the latter in the elastic metamaterials.  Enhanced asymptotic method of homogenization has been developed  to  provide the weak nonlocal effects as a small correction to  the local behavior \cite{boutin2012-nonlocal}. An approach has been presented \cite{willis2} for random elastic composites based on ensemble averaging of the material responses to a body force giving rise to  effective parameters of the medium depending on frequency and wavenumber. By this method the case of periodic media can be treated as well.

The nonlocal theory we use here,  takes fully the temporal dispersion and  spatial dispersion into account. The medium is assumed unbounded and homogeneous in the stationnary random statistical sense, therefore, the spatial dispersion refers only to the dependence of the permittivities, i.e. effective density and bulk modulus, on the Fourier wavenumbers $\bm{k}$ present in the macroscopic fields \cite{lanEL}.  The theory applies with some care to a periodic medium; in particular it gives the Bloch wavenumbers and defines Bloch impedances: indeed, it applies exactly to the ensemble defined by the random translation of one periodic sample.  The materials susceptible to show the nonlocal behavior may be classified into two main groups regarding their microgeometry. The first comprises the materials which exhibit this behavior in sufficiently high frequency regime. The second one concerns materials with microgeometry constituting the resonators, which exhibit spatial dispersion phenomena even at not very high frequencies; the resonance phenomena act as a source generating nonlocal behavior. In this article we investigate the second type of these geometries in the form of daisy chained Helmholtz resonators.  We will see the first one in a forthcoming  paper, where 1D propagation in a two-dimensional lattice of rigid cylinders will be studied.  A material made of an array of Helmholtz resonators filled by water has been studied experimentally, which has been found to show negative bulk modulus in the resonance frequency range \cite{fang}. Later, Helmhotz resonators as structural units were used to designe novel metamaterials  for  focusing ultrasound waves \cite{nick-lens} and broadband acoustic cloaking \cite{nick-cloak}.

Here, we will apply the nonlocal theory to quantatively describe the macroscopic dynamics of such a metamaterial filled with air as a viscothermal fluid, in 2D as well as in 3D. For the 2D case, using a simplified analytical solution of the complete equations, we will present how to obtain the nonlocal efffective density and effective bulk modulus. When these effective parameters  satisfy the dispersion equation based on  the nonlocal theory, we can compute the wavenumber of the least attenuated mode, among other modes. We can then check that the wavenumber resulting from the macroscopic nonlocal theory coincides with the Bloch wavenumber propagating and attenuating in the medium. The Bloch solution is determined using the same simplifying solving as in the nonlocal modeling, thus the results based on the two calculations should be comparable. Finally, as a check of the validity of the simplifying assumptions introduced in our modeling calculations, we have performed direct Finite Element Method (FEM) computations based on the exact equations in the framework of nonlocal homogenization.

In section \ref{general}, we review briefly the general framework of the nonlocal theory which will be used afterwards. The microscopic equations governing sound propagation in a rigid porous medium will be recalled, before mentioning the macroscopic Maxwellian equations describing the macroscopic nonlocal dynamics of the homogenized equivalent fluid.    In section \ref{nmodeling}, we will see the nonlocal modeling allowing the calculation of the effective parameters and the wavenumber of the least attenuated wave in the medium. The direct calculation of the Bloch wavenumber, using the similar simplifications, is presented in section \ref{bmodeling}. Section \ref{results} is devoted to the results of the three different calculations in 2D, and also the results based on the nonlocal modelling and Bloch wave calculations which have been generalized to 3D structures.

\section{Genereral framework of the nonlocal theory}\label{general}

In the following, we state the microscopic equations applied at the pore level, and the nonlocal Maxwellian macroscopic equations that describe the dynamics of the material as a homogeneous equivalent fluid medium. Then, we recall briefly the upscaling procedures allowing to obtain the frequency and wavenumber dependent effective parameters of the macroscopic equivalent fluid medium, i.e. effective density and effective bulk modulus \cite{lafarge2013}.

\subsection{Microscopic equations}\label{micro}

The dynamics of a small amplitude perturbation in a rigid-framed porous material filled with a viscothermal fluid is governed by the linearized equations of the mass, momentum, and energy balance, and a general fluid state equation as follows: in the fluid region $\mathcal{V}^{f}$

\begin{subequations}\label{viscth}
\begin{eqnarray}
\label{Navier}
\lefteqn{\rho_{0}\frac{\partial\bm{v}}{\partial t}=-\bm{\nabla} p +\eta\bm{\nabla}^2\bm{v}+(\zeta+\frac{\eta}{3})\bm{\nabla}(\bm{\nabla}\cdot\bm{v})}\\
\label{max1}
\lefteqn{\frac{\partial b}{\partial t} +\bm{\nabla}\cdot\bm{v}=0}\\
\label{statepb}
\lefteqn{\gamma\chi_{0}p=b+\beta_{0}\tau}\\
\label{Fourier}
\lefteqn{\rho_{0}c_{p}\frac{\partial \tau}{\partial t}=\beta_{0}T_{0}\frac{\partial p}{\partial t}+\kappa\bm{\nabla}^2\tau}
\end{eqnarray}
\end{subequations}
with boundary conditions 
\begin{eqnarray} \label{pore level bc}
\bm{v}=0, \ \ \
\tau = 0
\end{eqnarray}
applied to the fluid/solid interface $\partial \mathcal{V}$, where $\bm{v}$, $b\equiv\rho/\rho_{0}$, $p$ and $\tau$, are the fluid velocity, excess condensation, thermodynamic excess pressure, excess temperature, respectively, and $\rho$ is the excess density. The fluid constants $\rho_{0}$, $\eta$, $\zeta$, $\gamma$, $\chi_{0}$, $\beta_{0}$, $c_{p}$, $T_{0}$, $\kappa$, represent the ambient density, first viscosity, second viscosity, ratio of heat coefficients at constant pressure to constant volume $c_{p}/c_{v}$, adiabatic compressibility, coefficient of thermal expansion, specific heat coefficient at constant pressure, ambient temperature, and coefficient of thermal conduction, respectively.


\subsection{Macroscopic Maxwellian acoustics}\label{macro}

Before going through the macroscopic equations for sound propagation in rigid-framed porous media, and the homogenization procedure, we will  precise the notion of field averaging in the nonlocal approach.

\textbf{Averaging:} the present macroscopic theory is statistical in nature and has been developed in principle for fluid-saturated rigid-framed media which are homogeneous in an ensemble-averaged sense; this is the case of stationnary random media. The macroscopic properties represented in the theory refer to the ensemble of realizations. Thus for example, the propagation constants of the medium would refer to the propagation constant of \textit{coherent waves} in multiple-scattering theory. Here, the material we wish to study is not defined by stationary random realizations. It belongs on the contrary to the important class of periodic materials. The macroscopic theory can still be applied, however, if we consider the ensemble obtained by random translation of one sample. It turns out that the ensemble-average $\left\langle \; \right\rangle $ properties of the space are, in this case, precisely computable by spatial averaging over a periodic cell in a single realization. This, in a sense, reminds of ergodicity in the stationary random case.

The macroscopic condensation and velocity are defined as the average of pore scale microscopic fields: $\bm{V}\equiv\left\langle \bm{v}\right\rangle $, and  $B\equiv\left\langle b \right\rangle $; average over the periodic cell in the case of the periodic media. The electromagnetic analogy  suggests that the system of macroscopic equations can be carried through by introducing new Maxwellian fields $H$ and $\bm{D}$, and also operators $\hat{\rho}$ and $\hat{\chi}^{-1}$, such that

\begin{equation}\label{maxwell1}
\hspace*{-0.7cm}\text{\textbf{Field equations:}}\hspace*{1.7cm}\dfrac{\partial B}{\partial t}+\bm{\nabla} \cdot \bm{V}=0, \hspace*{1cm} \dfrac{\partial \bm{D}}{\partial t}=-\bm{\nabla} H
\end{equation}

\begin{equation}\label{maxwell2}
\hspace*{-0.7cm}\text{\textbf{Constitutive relations:}}\;\;\;\;\bm{D}=\hat{\rho}\bm{V}, \hspace*{2.3cm} H=\hat{\chi}^{-1}B
\end{equation}

where the integral operators of density $\hat{\rho}$ and bulk modulus $\hat{\chi}^{-1}$ are defined by 
\begin{subequations}\label{operator}
\begin{eqnarray}
\label{ma2}
\lefteqn{\bm{D}(t,\bm{r})=\int_{-\infty}^{t}dt'\int d\bm{r}' \rho(t-t',\bm{r}-\bm{r}')\bm{V}(t',\bm{r}')}\\ 
\label{ma3}
\lefteqn{H(t,\bm{r})=\int_{-\infty}^{t}dt'\int d\bm{r}' \chi^{-1}(t-t',\bm{r}-\bm{r}')B(t',\bm{r}')}
\end{eqnarray}
\end{subequations}

We notice that the kernels $\rho$ and $\chi^{-1}$ depend on the difference $t-t'$ and $\bm{r}-\bm{r}'$, which is due to the homogeneity in time and material space. That is why we can write (\ref{ma2}) and (\ref{ma3}) in the Fourier space, respectively, as
\begin{equation}\label{fourier}
\bm{D}(\omega,\bm{k})=\rho(\omega,\bm{k})\bm{V}(\omega,\bm{k}), \hspace*{2cm} H(\omega,\bm{k})=\chi^{-1}(\omega,\bm{k})B(\omega,\bm{k})
\end{equation}

In nonlocal theory,  the macroscopic $H$ field is defined through the Poynting-Schoch condition of \textit{acoustic part of energy current density} \cite{lafarge2013,nemati2014}:
\begin{equation}\label{pointing}
 S= H \bm{V}=\langle p \bm{v} \rangle
\end{equation}

Regarding  Eqs.(\ref{operator}) and (\ref{fourier}), it is visible that the theory allows for both temporal dispersion, shown by integration over time variable $t'$ in physical space and frequency dependence in Fourier space, and spatial dispersion, shown by  integration over space coordinates $\bm{r}'$ and wavenumber dependence in Fourier space. We will recognize the quantities in physical space $(t,\bm{r})$ and Fourier space $(\omega,\bm{k})$ by their arguments. Now, in order to clarify the relationship between constitutive operators and microgeometry, the kernel functions $\rho(\omega,\bm{k})$ and $\chi^{-1}(\omega,\bm{k})$ are needed to be determined, by introducing the procedures coarse-graining the dissipative fluid dynamics of the pore scale.


\subsection{Procedures to compute effective density and bulk modulus}\label{procedure}

In the 1D case of macroscopic propagation along a symmetry axis, for instance $x$-axis with the unit vector $\hat{\bm{x}}$, we will have $\bm{D}=D\hat{\bm{x}}$ and  $\bm{V}=V\hat{\bm{x}}$, $\bm{r}=x\hat{\bm{x}}$, and $\bm{k}=k\hat{\bm{x}}$ in the above equations (\ref{maxwell1}-\ref{pointing}).  To determine the Fourier functions $\rho(\omega,k)$ and $\chi^{-1}(\omega,k)$ for the 1D acoustic propagation in a medium with porosity $\phi$, we solve two independent action-response problems. For computing the effective density we consider the macroscopic response of the fluid subject to a single-component $(\omega,k)$ Fourier bulk force. The effective bulk modulus is related to the response of the fluid subject to a single-component Fourier rate of heat supply. 




\textbf{Two sets of equations to be solved:} the two systems of equations to be solved are written as

In the fluid region $\mathcal{V}_{f}$:
\begin{subequations}\label{action-response}
\begin{eqnarray}
\lefteqn{\dfrac{\partial b}{\partial t} + \nablam \cdot \bm{v} =0}\label{mass balance}\\
\lefteqn{\rho_{0}\dfrac{\partial \bm{v}}{\partial t}=-\nablam p +\eta \bm{\nabla}^{2}\bm{v}+\left(\zeta +\frac{1}{3}\eta\right)\nablam \left(\nablam  \cdot \bm{v}\right)+\underset{\text{Added for determination of density }}{\underbrace{\bm{F}e^{-i\omega t+ikx}  }}}\label{navier-stokes}\\
\lefteqn{\rho_{0}c_{p}\dfrac{\partial \tau }{\partial t}= \beta_{0} T_{0}\dfrac{\partial p}{\partial t}+\kappa \nablam^{2} \tau+\underset{\text{Added for determination of bulk modulus }}{\underbrace{ \dot{Q}e^{-i\omega t+ikx}}}}\label{energy blance}\\
\lefteqn{\gamma \chi_{0} p =b +\beta_{0} \tau}
\end{eqnarray}
\end{subequations}
\text{On the fluid/solid interface}$\; \partial \mathcal{V}$:
\begin{equation}\label{boundary}
\bm{v} =0, \hspace*{1cm} \tau = 0
\end{equation}
For convenience the excitation amplitudes are written as: $\dot{Q}e^{-i\omega t+ikx}=\beta_{0}T_{0} (\partial/\partial t)\left( \mathcal{P} e^{-i\omega t+ikx} \right) $,  and $\bm{F}e^{-i\omega t+ikx}=-\bm{\nabla}\left( \mathcal{P} e^{-i\omega t+ikx}\right)$.  Here, it is important to note that the excitation variables $\omega$ and $k$ are set as independent variables. The  solutions to the above systems  for the fields $p$, $b$, $\tau$, and components of $\bm{v}$ take the form $p(t,\bm{r})=p(\omega,k,\bm{r})e^{-i\omega t +ik x}$, and so on. Recall that the theory is formulated for a  geometry that is stationary random, and the averaging $\left\langle \; \right\rangle  $ is  the ensemble average. Thus here, the amplitude fields $\bm{v}(\omega,k,\bm{r})$,  $p(\omega,k,\bm{r})$, $b(\omega,k,\bm{r})$, and $\tau(\omega,k,\bm{r})$, are  stationary random functions of $\bm{r}$. Passing to the case of periodic geometry, we can limit ourselves to considering one periodic sample. The fields become periodic functions over a cell, and $\left\langle \; \right\rangle  $ is interpreted as a volume average over a cell. 

\textbf{Effective density and bulk modulus:} once the two systems of equations are solved independently, using the right hand  Maxwellian macroscopic equations in (\ref{maxwell1}) and (\ref{maxwell2}), we arrive at the following expressions for the nonlocal effective density and bulk modulus
\begin{subequations}
\begin{eqnarray}
\lefteqn{\rho(\omega,k)=\dfrac{k\left(\mathcal{P}+P(\omega,k) \right) }{\omega\left\langle v(\omega,k,\bm{r}) \right\rangle }}\label{rho}\\
\lefteqn{\chi^{-1}(\omega,k)=\dfrac{P(\omega,k)+\mathcal{P}}{\left\langle b(\omega,k,\bm{r})\right\rangle  +\phi\gamma\chi_{0}\mathcal{P}}}\label{chi}
\end{eqnarray}
\end{subequations}
where $P\langle\bm{v}\rangle =\langle p \bm{v}\rangle$, which has been inspired by (\ref{pointing}).

\textbf{Wavenumbers:} contrary to the case of local theory, here, since we fully take into account spatial dispersion, several normal mode solutions might exist, with fields varying as $e^{-i\omega t + ik x}$. Each solution should satisfy the following dispersion equation 
\begin{equation}\label{disp-nonlocal}
\rho(\omega,k)\chi(\omega,k)\omega^2=k^2
\end{equation}
which is easily extracted from the Maxwellian macroscopic equations. With each frequency $\omega$, several complex wavenumbers $k_{l}(\omega)$,  $\Im(k_{l})>0$, $l=1,2,...$, may be associated. 

In what follows, with the aim of obtaining the nonlocal effective density, effective bulk modulus, and wavenumber of the least attenuated mode, we will apply  this theoretical framework in analytical  simplified manner, to a 2D array of Helmholtz resonators, illustrated in Fig. \ref{geometry} right, exhibiting resonance phenomena which result in metamaterial behavior.


\section{Nonlocal modeling for 2D structure}
\label{nmodeling}

We proceed to determine the functions $\rho(\omega,k)$ and $\chi^{-1}(\omega,k)$ sufficiently precise to give an appropriate modeling of the least attenuated mode, which results then in purely frequency dependent functions $\rho(\omega)$ and $\chi^{-1}(\omega)$. To this aim, we need not consider in full detail the microscopic fields $\bm{v}$ and $p$.  In the waveguide $t$ and cavity $c$, instead of the microscopic fields, we can use the mean values $V_{t(c)}=\langle\bm{v}\rangle_{S}\cdot\hat{\bm{x}}$ and $P_{t(c)}=\langle p \rangle_{S}$, where $\langle\;\rangle_{S}$ denotes the average at a given $x$ over the waveguide or the cavity  width; and in the neck $n$, we can use the mean values $V_{n}=\langle\bm{v}\rangle_{S}\cdot\hat{\bm{y}}$ and $P_{n}=\langle p\rangle_{S}$, where  $\langle\;\rangle_{S}$ denotes the average at a given $y$ over the neck width, and $\hat{\bm{y}}$ is the unit vector in the $y$ direction. At the same time, we make some simplifications consistent with describing the propagation of these averaged quantities in terms of the Zwikker and Kosten  densities $\rho(\omega)$ and bulk modulii $\chi^{-1}(\omega)$, in the different slit portions. These depend only on the slit half-widths, which we shall denote by $s_t$, $s_n$, and $s_c$, in the tube, neck, and cavity. The different slit-like tube portions are illustrated in Fig.\ref{amplitude}. The main tube $t$ is divided in two Zwikker and Kosten ducts, a left duct, and a right duct, oriented in the $x$ direction. The same separation is made for the cavity $c$,  whereas the neck $n$ is not divided but seen as one Zwikker and Kosten duct oriented in $y$ direction.


\subsection{Determination of nonlocal effective density}
\label{nonlocal-modeling-rho}

Considering the periodic cell of Fig.\ref{geometry} right, and the corresponding cell average operation $\langle \; \rangle$, we look for the response of the fluid when a harmonic driving force $f(t,x)=fe^{-i \omega t +i kx}$ in the direction of $\bm{e}_{x}$ is applied. If we can determine the microscopic response velocity and pressure fields  $\bm{v}$, $p$, 
then we will have the function $\rho(\omega,k)$ through  the relation (see Eq.(\ref{rho}))
\begin{equation}\label{rho-res}
\rho(\omega,k)=\frac{f-ik\mathtt{P(\omega,k)}}{-i\omega \langle v(\omega,k,\bm{r}) \rangle}
\end{equation}
with $\mathtt{P}(\omega,k)=\left\langle p v\right\rangle /\left\langle v\right\rangle $, where the $v$ is the $x$-component of the microscopic velocity $\bm{v}$.

In \cite[Appendix]{nemati2014}, the Zwikker and Kosten local theory is expressed for tubes of circular cross-section. For 2D slits, exactly the same general principles of modeling may be used; only some details of the calculations are changed. 
In particular, the Bessel functions $J_0$ and $J_1$ are replaced by $\cosh$ and $\sinh$ functions. Zwikker and Kosten's effective densities $\rho_\alpha(\omega)$ and bulk modulii $\chi^{-1}_\alpha(\omega)$ in the guide, neck and cavity, will be \cite{allard}
\begin{figure}[h!]
\begin{center}
\includegraphics[width=7.5cm,angle=0]{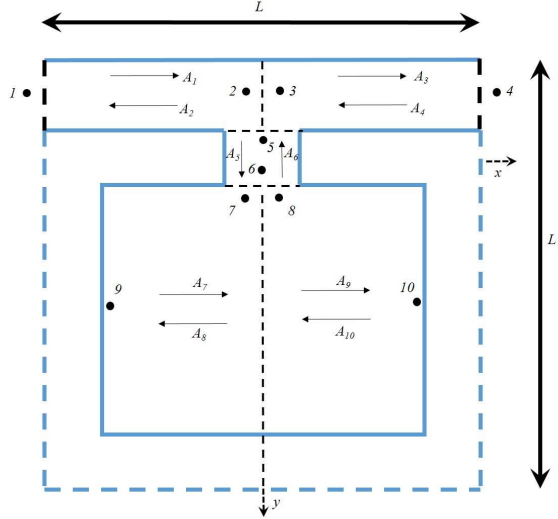} 
\end{center}
\caption{Illustration of slit portions and plane waves propagating in different parts of the resonator. Different positions are indicated by $m$, and different amplitudes by $A_{m}$, $m = 1, ..., 10$.}
\label{amplitude}
\end{figure}
\begin{equation}\label{slit}
\rho_{\alpha}(\omega)=\rho_{0}\left[ 1-\frac{\tanh \left( \sqrt{-i\omega \rho_{0}s_{\alpha}^{2}/\eta}\right) }{\sqrt{-i\omega \rho_{0}s_{\alpha}^{2}/\eta}}\right]^{-1},  \chi^{-1}_{\alpha}(\omega)=\gamma P_{0}\left[ 1+(\gamma-1) \frac{\tanh \left( \sqrt{-i\omega \rho_{0}c_{p}s_{\alpha}^{2}/\kappa}\right) }{\sqrt{-i\omega \rho_{0}c_{p}s_{\alpha}^{2}/\kappa}}\right]^{-1}
\end{equation}
for $\alpha =t,n,c$, where the indexes $t$, $n$, and $c$ are related to the tube, neck, and cavity respectively,  $\rho_{0}$, and $P_{0}$ the fluid pressure at rest. The corresponding wavenumbers $k_{\alpha}(\omega)$ and characteristic admittances $Y_{\alpha}(\omega)$ are expressed as $k_{\alpha}=\omega/c_{\alpha}$, and $Y_{\alpha}(\omega)=2s_{\alpha}/(\rho_{\alpha}c_{\alpha})$, for $\alpha =t,n,c$, where $c_{\alpha}=1/\sqrt{\rho_\alpha \chi_\alpha}$, is the corresponding Zwikker and Kosten's phase velocity. Notice that we include the slit width $2s_{\alpha}$ (resp. $\Sigma, \sigma$, and $L-\Sigma-2l$ in the resonator, see Fig. \ref{geometry}, left) in the definition of the characteristic admittance, because it simplifies the subsequent writing of continuity conditions.



We start writing the Zwikker and Kosten's equations in the different parts of the periodic cell.  For the tube and the cavity, i.e. , $\alpha=t, c$, we have 
\begin{subequations}\label{zwikker-nonlocal}
\begin{eqnarray} 
\lefteqn{- i \omega\frac{ \rho_{\alpha}(\omega)}{S_{\alpha}}V_{\alpha}=-\frac{\partial P_{\alpha}}{\partial x}+f e^{ikx}}\\
\lefteqn{i\omega S_{\alpha} \chi_{\alpha}(\omega)P_{\alpha}=\frac{\partial V_{\alpha}}{\partial x}}
\end{eqnarray}
\end{subequations}
where, $V_{\alpha}=V_{x}S_{\alpha}$ is the flow rate field across the cross section $S_{\alpha}$, with $V_{x}$ the $x$-component of the velocity in the sense of Zwikker and Kosten (averaged over the section), and $P_{\alpha}$ is the Zwikker and Kosten's pressure. In the neck, the external excitation having no $y$-component 
\begin{subequations}\label{zwikker-res-neck}
\begin{eqnarray} 
\lefteqn{ i \omega\frac{ \rho_{n}(\omega)}{\sigma}V_{n}=\frac{\partial P_{n}}{\partial y}}\\
\lefteqn{i\omega \sigma \chi_{n}(\omega)P_{n}=\frac{\partial V_{n}}{\partial y}}
\end{eqnarray}
\end{subequations}
where, $V_{n}=V_{y}\sigma$ is the flow rate, with $V_{y}$ the $y$-component of the velocity, and $P_{n}$ is the Zwikker and Kosten's pressure in the neck. 
  
The general solution of the non homogeneous equations in the tube and the cavity, $(P_{\alpha}, V_{\alpha})$, $\alpha=t, c$, is written as the sum of the general solution $(P_{\alpha,h},V_{\alpha,h})$ of the homogeneous equations and a particular solution $(P_{\alpha,p}, V_{\alpha,p})$ of the non homogeneous equations. A general solution of the homogeneous equations (\ref{zwikker-nonlocal}) is written as
\begin{equation}\label{homog-solution-tube}
\begin{pmatrix}
P_{\alpha,h}\\
V_{\alpha,h}
\end{pmatrix}
=\begin{pmatrix}
1\\
Y_{\alpha}
\end{pmatrix}
A^{+}e^{ik_{\alpha}x}+
\begin{pmatrix}
1\\
-Y_{\alpha}
\end{pmatrix}
A^{-}e^{-ik_{\alpha}x}
\end{equation}
where $A^{+}$ and $A^{-}$ are the amplitudes of the plane waves in direction of the positive $x$-axis and  negative $x$-axis, respectively. The following particular solution can be considered
\begin{equation}\label{particular-solution-tube-cavity}
\begin{pmatrix}
P_{\alpha,p}\\
V_{\alpha,p}
\end{pmatrix}
=
\begin{pmatrix}
B_{\alpha}\\
C_{\alpha}
\end{pmatrix}
fe^{ikx}
\end{equation}
where $B_{\alpha}$ and $C_{\alpha}$ are four constants (for each $\omega$) to be determined. Substituting (\ref{particular-solution-tube-cavity}) in (\ref{zwikker-nonlocal}) gives the four constants $B_{t}=ik/(\omega^{2}\rho_{t}\chi_{t}-k^{2})$, $C_{t}=i\omega \chi_{t}\Sigma/(\omega^{2}\rho_{t}\chi_{t}-k^{2})$, $B_{c}=ik /(\omega^{2}\rho_{c}\chi_{c}-k^{2})$, and $C_{c}= i\omega \chi_{c}(L-\Sigma-2l)/(\omega^{2}\rho_{c}\chi_{c}-k^{2})$. The particular solution is the same in the left and right portions of the tube and the cavity. On the contrary and because of the presence of the neck, the general solution will have different amplitude constants in the left and right portions. Thus, the general solution of Eqs.(\ref{zwikker-nonlocal}) can be written as
\begin{subequations}\label{general-tube-cavity}
\begin{eqnarray}\label{left-right-tube}
\lefteqn{\begin{pmatrix}
P_{t}\\
V_{t}
\end{pmatrix}
=
\begin{pmatrix}
1\\
Y_{t}
\end{pmatrix}
A_{1,3}fe^{ik_{t}x}+
\begin{pmatrix}
1\\
-Y_{t}
\end{pmatrix}
A_{2,4}f e^{-ik_{t}x}+
\begin{pmatrix}
B_{t}\\
C_{t}
\end{pmatrix}
f e^{ikx}}\\
\label{left-right-cavity}
\lefteqn{\begin{pmatrix}
P_{c}\\
V_{c}
\end{pmatrix}
=
\begin{pmatrix}
1\\
Y_{c}
\end{pmatrix}
A_{7,9}f e^{ik_{c}x}+
\begin{pmatrix}
1\\
-Y_{c}
\end{pmatrix}
A_{8,10}fe^{-ik_{c}x}+
\begin{pmatrix}
B_{c}\\
C_{c}
\end{pmatrix}
f e^{ikx}}
\end{eqnarray}
\end{subequations}
where  (\ref{left-right-tube}) with amplitudes $A_{1}$ and $A_{2}$ corresponds to the left part of the tube,  and with amplitudes $A_{3}$ and $A_{4}$ to the right part (Fig. \ref{amplitude}); similarly for (\ref{left-right-cavity}): $A_{7}$ and $A_{8}$ for the left part of the cavity, and $A_{9}$ and $A_{10}$ for the right part (Fig. \ref{amplitude}). These eight amplitudes are to be determined. The general solution of Eqs.(\ref{zwikker-res-neck}), $(P_{n},V_{n})$ has the form 
\begin{equation}\label{general-solution-neck}
\begin{pmatrix}
P_{n}\\
V_{n}
\end{pmatrix}
=
\begin{pmatrix}
1\\
Y_{n}
\end{pmatrix}
A_{5}f e^{ik_{n}y}+
\begin{pmatrix}
1\\
-Y_{n}
\end{pmatrix}
A_{6}f e^{-ik_{n}y}
\end{equation}
where $A_{5}$ and  $A_{6}$ are the neck amplitude-relating constants to be determined (Fig. \ref{amplitude}).    

Indeed, in the framework of our simple plane-wave modeling, there are $10$ relations concerning the flow rate and pressure, which are assumed to be verified. These continuity relations involve the values of the fields at different locations indicated by numbers $m=1,...,10$,  in Fig.\ref{amplitude}. We now proceed to write them.

The Bloch condition results in $P_{t}^{(4)}=e^{ikL}P_{t}^{(1)}$ and $V_{t}^{(4)}=e^{ikL}V_{t}^{(1)}$. Then $A_{3}e^{ik_{t}L/2}+A_{4}e^{-ik_{t}L/2}=e^{ikL} ( A_{1}e^{-ik_{t}L/2}+A_{2}e^{ik_{t}L/2})$ and  $A_{3}e^{ik_{t}L/2}-A_{4}e^{-ik_{t}L/2}=e^{ikL} ( A_{1}e^{-ik_{t}L/2}-A_{2}e^{ik_{t}L/2})$. We assume the continuity of the pressure at the junction (2)-(3),  $P_{t}^{(3)}=P_{t}^{(2)}$, then $A_{3}+A_{4}=A_{1}+A_{2}$. We assume the continuity of the pressure at the junction (5)-(2), $P_{n}^{(5)}=P_{t}^{(2)}$, then $A_{5}e^{-ik_{n}l/2}+A_{6}e^{ik_{n}l/2}=A_{1}+A_{2}+B_{t}$. The flow rate at the junction (2)-(3)-(5) is assumed to verify $V_{t}^{(2)}-V_{t}^{(3)}=V_{n}^{(5)}$, which yields $Y_{t}\left(A_{1}-A_{2}-A_{3}+A^{4} \right)=Y_{n}\left(A_{5}e^{-ik_{n}l/2}-A_{6}e^{ik_{n}l/2} \right)$. The continuity of the pressure at the junction (6)-(7), $P_{n}^{(6)}=P_{c}^{(7)}$ results in $A_{5}e^{ik_{n}l/2}+A_{6}e^{-ik_{n}l/2}=A_{7}+A_{8}+B_{c}$. The flow rate at the junction (6)-(7)-(8) is assumed to verify $V_{n}^{(6)}+V_{c}^{(7)}=V_{c}^{(8)}$, $Y_{n}\left(A_{5}e^{ik_{n}l/2}-A_{6}e^{-ik_{n}l/2} \right) +Y_{c}(A_{7}-A_{8})=Y_{c}(A_{9}-A_{10})$. The pressure is continuous at (7)-(8), $P_{c}^{(7)}=P_{c}^{(8)}$ then, $A_{7}+A_{8}=A_{9}+A_{10}$. The flow rate vanishes at the interface solid-fluid,  $V_{c}^{(9)}=0$, we have $Y_{c}\left[ A_{7}e^{-ik_{c}(L-l)/2}-A_{8}e^{ik_{c}(L-l)/2}\right]=-C_{c}e^{-ik(L-l)/2}$. The flow rate vanishes at the interface solid fluid, $V_{c}^{(10)}=0$, we have $Y_{c}\left[ A_{9}e^{ik_{c}(L-l)/2}-A_{10}e^{-ik_{c}(L-l)/2}\right]=-C_{c}e^{ik(L-l)/2}$.

As such, we have $10$ equations  for $10$ unknown amplitudes $A_{1},..,A_{10}$. Once these are determined, we will have all the Zwikker and Kosten's fields through  Eqs.(\ref{left-right-tube}), (\ref{general-solution-neck}), and (\ref{left-right-cavity}). At this point, we can easily obtain the cell averages $\left\langle v \right\rangle $ and $\left\langle pv \right\rangle $. Let us start with $\left\langle v \right\rangle $ regarding the fact that the Zwikker and Kosten's flow rate has no component along the $x$-axis
\begin{equation}\label{average-v1-rho}
\left\langle v \right\rangle = \frac{1}{L^{2}}\left(\int_{-L/2}^{0} V_{t}\;dx+ \int_{0}^{L/2} V_{t}\;dx+ \int_{-(L-l)/2}^{0} V_{c}\;dx + \int_{0}^{(L-l)/2} V_{c}\;dx\right) 
\end{equation}
Similarly, we can compute $\left\langle pv \right\rangle$ through the following relation
\begin{equation}\label{average-pv1-rho}
\left\langle pv \right\rangle = \frac{1}{L^{2}}\left(\int_{-L/2}^{0} P_{t}V_{t}\;dx+ \int_{0}^{L/2} P_{t}V_{t}\;dx+ \int_{-(L-l)/2}^{0} P_{c}V_{c}\;dx + \int_{0}^{(L-l)/2} P_{c}V_{c}\;dx\right) 
\end{equation}

Now, we can obtain explicitly the effective density function $\rho(\omega,k)$ through Eq.(\ref{rho-res}). In the next section, the effective bulk modulus is computed in a similar way but with a different excitation term, and with exactly the same conditions on the flow rate and pressure fields at different junctions.


\subsection{Determination of nonlocal effective bulk modulus}
\label{nonlocal-modeling-chi}

Considering the periodic cell  (Fig.\ref{amplitude}), when a harmonic  heating $\dot{Q}(t,x)=\dot{Q}_{0}e^{-i \omega t +i kx}=-i\omega\beta_0 T_0 \mathcal{P}e^{-i \omega t +i kx}$  is applied in the medium,  we write the Zwikker and Kosten's equations, in each  part of the resonator: tube, neck, and cavity. 
The aim is to obtain the function $\chi^{-1}(\omega,k)$ as it is indicated in Eq.(\ref{chi}).
In the main tube and the cavity, for $\alpha=t, c$, we write 
\begin{subequations}\label{zwikker-tube-cavity-chi}
\begin{eqnarray} 
\lefteqn{- i \omega\frac{ \rho_{\alpha}(\omega)}{S_{\alpha}}V_{\alpha}=-\frac{\partial P_{\alpha}}{\partial x}}\\
\lefteqn{i\omega S_{\alpha} \chi_{\alpha}(\omega)P_{\alpha}+i\omega S_{\alpha}\left( \chi_{\alpha}(\omega)-\gamma \chi_{0}\right)\mathcal{P} =\frac{\partial V_{\alpha}}{\partial x}}
\end{eqnarray}
\end{subequations}
The second term in the second equation might not seem to be obvious but follows  the very procedure of obtaining (\ref{chi}).
In the neck, the equations are written as 
\begin{subequations}\label{zwikker-neck-chi}
\begin{eqnarray} 
\lefteqn{ i \omega\frac{ \rho_{n}(\omega)}{\sigma}V_{n}=\frac{\partial P_{n}}{\partial y}}\\
\lefteqn{i\omega \sigma \chi_{n}(\omega)P_{n}+i\omega \sigma \left(  \chi_{n}(\omega)-\gamma \chi_{0}\right)\mathcal{P} \left\langle e^{ikx}\right\rangle _{\sigma} =\frac{\partial V_{n}}{\partial y}}
\end{eqnarray}
\end{subequations}
where the term $\mathcal{P} \left\langle e^{ikx}\right\rangle _{\sigma}$ comes from the averaging of $\dot{Q}$ over the neck section. Here also, the second equation  might not appear obvious, but follows the procedure of the determination (\ref{chi})  seen in nonlocal theory. 

As in the section \ref{nonlocal-modeling-rho}, the general solution of the non homogeneous equations (\ref{zwikker-tube-cavity-chi}) in the right or left part of the tube and the cavity, is written as the sum of the general solution $(P_{\alpha,h},V_{\alpha,h})$ of the homogeneous equations and a particular solution $(P_{\alpha,p}, V_{\alpha,p})$ of the  non homogeneous equations. A general solution of the homogeneous equations (\ref{zwikker-tube-cavity-chi}) is written as Eq.(\ref{homog-solution-tube}). The following particular solution can be considered
\begin{equation}\label{particular-solution-tube-chi}
\begin{pmatrix}
P_{\alpha,p}\\
V_{\alpha,p}
\end{pmatrix}
=
\begin{pmatrix}
B_{\alpha}\\
C_{\alpha}
\end{pmatrix}
\mathcal{P}e^{ikx}
\end{equation}
where $B_{\alpha}$ and $C_{\alpha}$ are constants to be determined. Substituting (\ref{particular-solution-tube-chi}) in (\ref{zwikker-tube-cavity-chi}) gives the four constants $B_{t}=\omega^{2}\rho_{t}(\chi_{t}-\gamma \chi_{0})/(k^{2}-\omega^{2}\rho_{t}\chi_{t})$, $C_{t}=\omega k (\chi_{t}-\gamma \chi_{0})\Sigma/(k^{2}-\omega^{2}\rho_{t}\chi_{t})$, $B_{c}=\omega^{2}\rho_{c}(\chi_{c}-\gamma \chi_{0})/(k^{2}-\omega^{2}\rho_{c}\chi_{c})$, and $C_{c}=\omega k (\chi_{t}-\gamma \chi_{0})(L-\Sigma -2L)/ (k^{2}-\omega^{2}\rho_{c}\chi_{c}) $. Thus, the general solution Eqs.(\ref{zwikker-tube-cavity-chi}) can be written as Eqs.(\ref{general-tube-cavity}), replacing $f$ with $\mathcal{P}$. The amplitudes $A_{1}$, $A_{2}$, $A_{3}$, $A_{4}$, $A_{7}$, $A_{8}$, $A_{9}$, and $A_{10}$ (Fig. \ref{amplitude}) are to be determined. 

As for the tube and the cavity, the general solution of the non homogeneous equations (\ref{zwikker-neck-chi}) in the neck, is written as the sum of the general solution $(P_{n,h},V_{n,h})$ of the homogeneous equations  and a particular solution $(P_{n,p}, V_{n,p})$ of the non homogeneous equations.  We can find a particular solution in the following form
\begin{equation}\label{particular-solution-neck-chi}
\begin{pmatrix}
P_{n,p}\\
V_{n,p}
\end{pmatrix}
=
\begin{pmatrix}
B_{n}\\
C_{n}
\end{pmatrix}
\mathcal{P}
\end{equation}
where $B_{n}$ and $C_{n}$ are two constants which will be determined by substituting (\ref{particular-solution-neck-chi}) in (\ref{zwikker-neck-chi}):\\ $B_{n}=(2/k\sigma) \left(\gamma \chi_{0}/\chi_{n}-1\right) \sin(k\sigma/2)$, and $C_{n}=0$. To obtain the above expression for $B_{n}$, the average $\left\langle e^{ikx}\right\rangle _{\sigma}$  has been easily calculated 
\begin{equation*}
 \left\langle e^{ikx}\right\rangle _{\sigma} =\frac{1}{\sigma}\int_{-\sigma /2}^{\sigma /2} e^{ikx}dx=\frac{2}{k\sigma}\sin\left( \frac{k\sigma}{2} \right) 
\end{equation*}
Thus, the general solution of Eq.(\ref{zwikker-neck-chi}) in the neck can be written as
\begin{equation}\label{general-solution-left-neck-chi}
\begin{pmatrix}
P_{n}\\
V_{n}
\end{pmatrix}
=
\begin{pmatrix}
1\\
Y_{n}
\end{pmatrix}
A_{5}\mathcal{P}e^{ik_{n}y}+
\begin{pmatrix}
1\\
-Y_{c}
\end{pmatrix}
A_{6}\mathcal{P}e^{-ik_{n}y}+
\begin{pmatrix}
B_{n}\\
0
\end{pmatrix}
\mathcal{P}\\
\end{equation}
where  $A_{5}$ and $A_{6}$ are amplitude-relating constants to be determined (Fig. \ref{amplitude}). 

As in the previous section \ref{nonlocal-modeling-rho}, in the framework of our modeling, there are $10$ relations which are assumed to be verified,  allowing to relate the flow rate and pressures at different indicated points in Fig.\ref{amplitude}. These relations result in $10$ equations by which we can compute the amplitudes $A_{1}$, ..., $A_{10}$. Consequently, all Zwikker and Kosten's fields will be found.  The averages $\left\langle v \right\rangle $ and $\left\langle pv \right\rangle $ are found through rewriting the equations (\ref{average-v1-rho}) and (\ref{average-pv1-rho}) for the actual fields. We need also the expression for $\left\langle b \right\rangle $ to obtain $\chi^{-1}(\omega,k)$. We have
\begin{eqnarray*}\label{average-b}
-i\omega \left\langle b\right\rangle &=& -\frac{1}{L^{2}}\int \bm{\nabla} \cdot \bm{v}\;dx dy\\ \nonumber
&=&  -\frac{1}{L^{2}} \oint \bm{v}\cdot \bm{n} \;dS=  -\frac{1}{L^{2}} \left(-V_{t}^{(1)}+V_{t}^{(4)}\right)\\ \nonumber
&=&-\frac{\mathcal{P}}{L^{2}} \left[ 2iC_{t}\sin\frac{k_{t}L}{2}+Y_{t}\left(  -A_{1}e^{-i\frac{k_{t}L}{2}}+ A_{2}e^{i\frac{k_{t}L}{2}}+A_{3}e^{i\frac{k_{t}L}{2}}-A_{4}e^{-i\frac{k_{t}L}{2}}\right)  \right] 
\end{eqnarray*}
where $\bm{n}$ is the normal unit vector outward from the border of integration.  

Now, we can obtain explicitly the effective bulk modulus function $\chi^{-1}(\omega,k)$ through Eq.(\ref{chi}).


\section{Bloch wave modeling}
\label{bmodeling}

In this section we directly seek, without using the principles of the nonlocal macroscopic theory but within the same plane wave modeling, the macroscopic Bloch wavenumber $k_{B}$ of the least attenuated wave propagating in the direction of positive $x$-axis, such that 

\begin{equation}\label{bloch}
\begin{pmatrix}
P^{(4)}_{t}\\
V^{(4)}_{t}
\end{pmatrix}
=
e^{ik_{B}L}
\begin{pmatrix}
P^{(1)}_{t}\\
V^{(1)}_{t}
\end{pmatrix}
\end{equation}  
To the field constituted of $10$ Zwikker and Kosten's slit waves, as illustrated in Fig.\ref{amplitude}, are associated $10$ complex amplitudes $A_{1},...,A_{10}$. 
As before, on these $10$ amplitudes there are $2$ relations (\ref{bloch}) expressing the Bloch condition, and  $8$ relations expressing the continuity equations. All these relations are now homogeneous relations, so that nontrivial solutions will be obtained only if the determinant vanishes. This condition will give the Bloch wavenumber $k_B$. 

The first step is to determine the entrance admittance of the resonator  $Y_{r}=V_{n}^{(5)}/P_{n}^{(5)}$. The general solution of the homogeneous form of Eqs.(\ref{zwikker-nonlocal}) for the cavity, $\alpha=c$, without the forcing term, is written as
\begin{eqnarray}\label{bloch-cavity}
\begin{pmatrix}
P_{c}\\
V_{c}
\end{pmatrix}
=
\begin{pmatrix}
1\\
Y_{c}
\end{pmatrix}
A_{7,9}e^{ik_{c}x}+
\begin{pmatrix}
1\\
-Y_{c}
\end{pmatrix}
A_{8,10}e^{-ik_{c}x}
\end{eqnarray}
where $A_{7}$ and $A_{8}$ are the amplitudes of the waves in the left part of the cavity, and $A_{9}$, $A_{10}$ are the amplitudes of the waves in the right part.  Regarding the above equation, the three conditions $P_{c}^{(7)}=P_{c}^{(8)}$, $V_{c}^{(9)}=0$, and $V_{c}^{(10)}=0$, result in the three following relations  $A_{8}=A_{7}e^{-ik_{c}(L-l)}$, $A_{9}=A_{7}e^{-ik_{c}(L-l)}$, and $A_{10}=A_{7}$. Using (\ref{bloch-cavity}) there follows $P_{c}^{(7)}=A_{7}\left(1-e^{ik_{c}(L-l)} \right)$, $V_{c}^{(7)}= Y_{c}A_{7}\left(1+e^{ik_{c}(L-l)} \right)$, and $V_{c}^{(8)}= Y_{c}A_{7}(e^{ik_{c}(L-l)}-1) $. Then, we can obtain the expressions for $P_{n}^{(6)}$ and $V_{n}^{(6)}$, through  already indicated continuity conditions $P_{n}^{(6)}=P_{c}^{(7)}$, and $V_{n}^{(6)}+V_{c}^{(7)}=V_{c}^{(8)}$, which, subsequently, yields the impedance $Y_{6}=V_{n}^{(6)}/P_{n}^{(6)}$ 
\begin{equation}\label{y6}
\lefteqn{Y_{6}=-2iY_{c}\dfrac{1-e^{-ik_{c}(L-l)}}{1+e^{-ik_{c}(L-l)}}}
\end{equation}

Once $P_{n}^{(6)}$ and $V_{n}^{(6)}$ are known, we can obtain $P_{n}^{(5)}$ and $V_{n}^{(5)}$ through
\begin{equation}
\begin{pmatrix}
P_{n}^{(5)}\\
V_{n}^{(5)}
\end{pmatrix}
=
\begin{pmatrix}
\cos k_{n}l &-\frac{i}{Y_{n}}\sin k_{n}l\\
-iY_{n}\sin k_{n}l & \cos k_{n}l
\end{pmatrix}
\begin{pmatrix}
P_{n}^{(6)}\\
V_{n}^{(6)}
\end{pmatrix}
\end{equation}

Thus, the impedance of the resonator $Y_{r}$ is expressed as 

\begin{equation}\label{yr}
Y_{r}=\frac{-iY_{n}\sin k_{n}l+Y_{6} \cos k_{n}l}{ \cos k_{n}l-i\dfrac{Y_{6}}{Y_{n}}\sin k_{n}l}
\end{equation}
Now, we look for the macroscopic wavenumber $k_{B}$. The following relations are satisfied in the right and left part of the tube
\begin{eqnarray}
\begin{pmatrix}
P_{t}^{(1),(3)}\\
V_{t}^{(1),(3)}
\end{pmatrix}
=
\begin{pmatrix}
\cos \frac{k_{t}L}{2} &-\frac{i}{Y_{t}}\sin \frac{k_{t}L}{2}\\
-iY_{t}\sin \frac{k_{t}L}{2} & \cos \frac{k_{t}L}{2}
\end{pmatrix}
\begin{pmatrix}
P_{t}^{(2),(4)}\\
V_{t}^{(2),(4)}
\end{pmatrix}
\end{eqnarray}
Making use of Eq.(\ref{bloch}), the above equations result in
\begin{equation}\label{3-et-2}
\begin{pmatrix}
P_{t}^{(3)}\\
V_{t}^{(3)}
\end{pmatrix}
= e^{ik_{B}L}
\begin{pmatrix}
\cos k_{t}L &-\frac{i}{Y_{t}}\sin k_{t}L\\
-iY_{t}\sin k_{t}L & \cos k_{t}L
\end{pmatrix}
\begin{pmatrix}
P_{t}^{(2)}\\
V_{t}^{(2)}
\end{pmatrix}
\end{equation}
On the other hand, as we have seen before, the three following conditions are assumed in the resonator: $P_{t}^{(3)}=P_{t}^{(2)}$, $P_{n}^{(5)}=P_{t}^{(2)}$, and $V_{t}^{(2)}-V_{t}^{(3)}=V_{n}^{(5)}$. We have immediately $P_{t}^{(3)}=P_{t}^{(2)}=(1/Y_{r})\left(V_{t}^{(2)}-V_{t}^{(3)} \right)$. Writing the two equations resulting from (\ref{3-et-2}), and eliminating $P_{t}^{(3)}$ and $P_{t}^{(2)}$ in these equations, gives
\begin{equation}
\begin{pmatrix}
\frac{1}{Y_{r}}- e^{ik_{B}L}\left(\frac{1}{Y_{r}}\cos k_{t}L-\frac{i}{Y_{t}}\sin k_{t}L \right) & -\frac{1}{Y_{r}}\left(1 -e^{ik_{B}L}\cos k_{t}L \right) \\
e^{ik_{B}L}\left(i\frac{Y_{t}}{Y_{r}}\sin k_{t}L-\cos k_{t}L \right) & 1-e^{ik_{B}L}\frac{iY_{t}}{Y_{r}}\sin k_{t}L
\end{pmatrix}
\begin{pmatrix}
V_{t}^{(2)}\\
V_{t}^{(3)}
\end{pmatrix}
=
\begin{pmatrix}
0\\
0
\end{pmatrix}
\end{equation}
The determinant of the coefficient matrix must vanish, if the above equations  have non-zero solutions. This  yields a second degree algebraic equation $e^{2ik_{B}L}- De^{ik_{B}L}+1=0$, with $D=2\cos k_{t}L-i (Y_{r}/Y_{t}) \sin k_{t}L  $. This gives immediately the Bloch wavenumber

\begin{equation}\label{kB}
k_{B}=-\frac{i}{L}\ln \left(\frac{D}{2}\pm \sqrt{\frac{D^{2}}{4}-1} \right) 
\end{equation}


\section{Results}\label{results}

Here, we present the results of the nonlocal modeling, Bloch wave modeling and FEM simulations for the two-dimensional metamaterial made by Helmholtz resonators. Once the  simplified nonlocal and Bloch wave modeling are validated by the results of the FEM simulations which are based on the solutions of the exact equations (\ref{action-response}), we employ the same nonlocal modeling framework to compute the macroscopic acoustic properties of the three-dimensional material. For both  2D and 3D structures, the resonators are filled with air as a viscothermal fluid. The fluid properties for all computations are indicated in Table \ref{constants}. In 2D and 3D cases, the results relating to the wavenumber of the least attenuated mode and the effective bulk modulus of the material will be shown, versus a frequency adimensional parameter. Moreover, we will present a simple method allowing to obtain the 2D geometry roughly equivalent to the 3D material, regarding the macroscopic dynamic behavior of the material in the resonance regime of the fundamental mode.

\begin{table}[h!]
\centering
\caption{Fluid properties used in all computations.}\label{constants}
\begin{tabular}{|c|c|c|c|c|c|c|c|c|}
\hline
$\rho_{0}$&
$T_{0}$&
$c_{0}$&
$\eta$&
$\zeta$&
$\kappa$&
$\chi_{0}$&
$c_{p}$&
$\gamma$\\
($kg/m^{3}$) & ($K$) & $(m/s)$ & $(kg\;ms^{-1})$ & ($kg\;ms^{-1}$) & ($Wm^{-1}K^{-1}$)& ($Pa^{-1}$) & ($J\;kg^{-1}K^{-1}$) & \\
\hline
\hline
1.205 &  293.5 & 340.14 & $1.84\times 10^{-5}$ & 0.6 $\eta$ & $2.57\times 10^{-2}$ & $7.17\times 10^{-6}$ & 997.54 & 1.4 \\
\hline
\end{tabular}
\end{table}

%


\subsection{2D structure filled with air}

For the geometry considered in Fig. \ref{geometry} right, to perform the computations, we have set $L=1cm$, $\Sigma=0.2L$, and $\sigma=0.015L$. The functions $\rho(\omega,k)$ and $\chi^{-1}(\omega,k)$ are first determined  within the approximations of our nonlocal modeling in section \ref{nmodeling}. Given these expressions, we know that according to nonlocal theory the possible wavenumbers in the medium will be the solutions of the dispersion relation (\ref{disp-nonlocal}). Solving the equation (\ref{disp-nonlocal}) by a Newton-Raphson scheme, we have checked that the obtained expressions for $\rho(\omega,k)$ and $\chi^{-1}(\omega,k)$ are such that a complex solution $k(\omega)$ to (\ref{disp-nonlocal}) exists, very close to the value $k_{B}(\omega)$ in (\ref{kB}). 
The frequency dependent effective density $\rho(\omega,k(\omega))=\rho(\omega)$, and effective bulk modulus $\chi^{-1}(\omega,k(\omega))=\chi^{-1}(\omega)$, are then obtained by putting $k=k(\omega)$ in the aforementioned excitation terms (sections \ref{nonlocal-modeling-rho} and \ref{nonlocal-modeling-chi}). 

Solving the equation (\ref{disp-nonlocal}) by the Newton-Raphson method, we varied frequency step by step, taking as initial value for $k(\omega)$ at a given frequency, the solution value obtained at the preceding frequency. Only for the starting frequency $\omega_{0}$ in the range of interest,  we have chosen the value $k_{B}(\omega_{0})$ with a $10\%$ discrepancy. 

To ascertain the validity of the modeling we have also performed direct FEM solving of the action-response problems, hence giving FEM evaluations of the functions $\rho(\omega,k)$ and  $\chi^{-1}(\omega,k)$. From these functions, the computation of the wavenumber of the least attenuated wave 
was performed in the same way as just seen, with the only difference that (for computation time reason) the initial $k(\omega)$ value at a given frequency was systematically taken to be $k_{B}(\omega)$ with $10\%$ discrepancy. Finally, FEM evaluations of the frequency dependent effective density $\rho(\omega,k(\omega))=\rho(\omega)$, and effective bulk modulus $\chi^{-1}(\omega,k(\omega))=\chi^{-1}(\omega)$, were obtained by putting $k=k(\omega)$ in the aforementioned excitation terms.



The FEM computations have been performed using FreeFem++ \cite{freefem}, an open source tool solving partial differential equations. Adaptive meshing was employed. According to all of the calculations, the effective density remains practically constant and, therefore, does not play an important role in the macroscopic dynamics of this material.

\begin{figure}[h!]
\begin{center}
\includegraphics[width=7.5cm,angle=0]{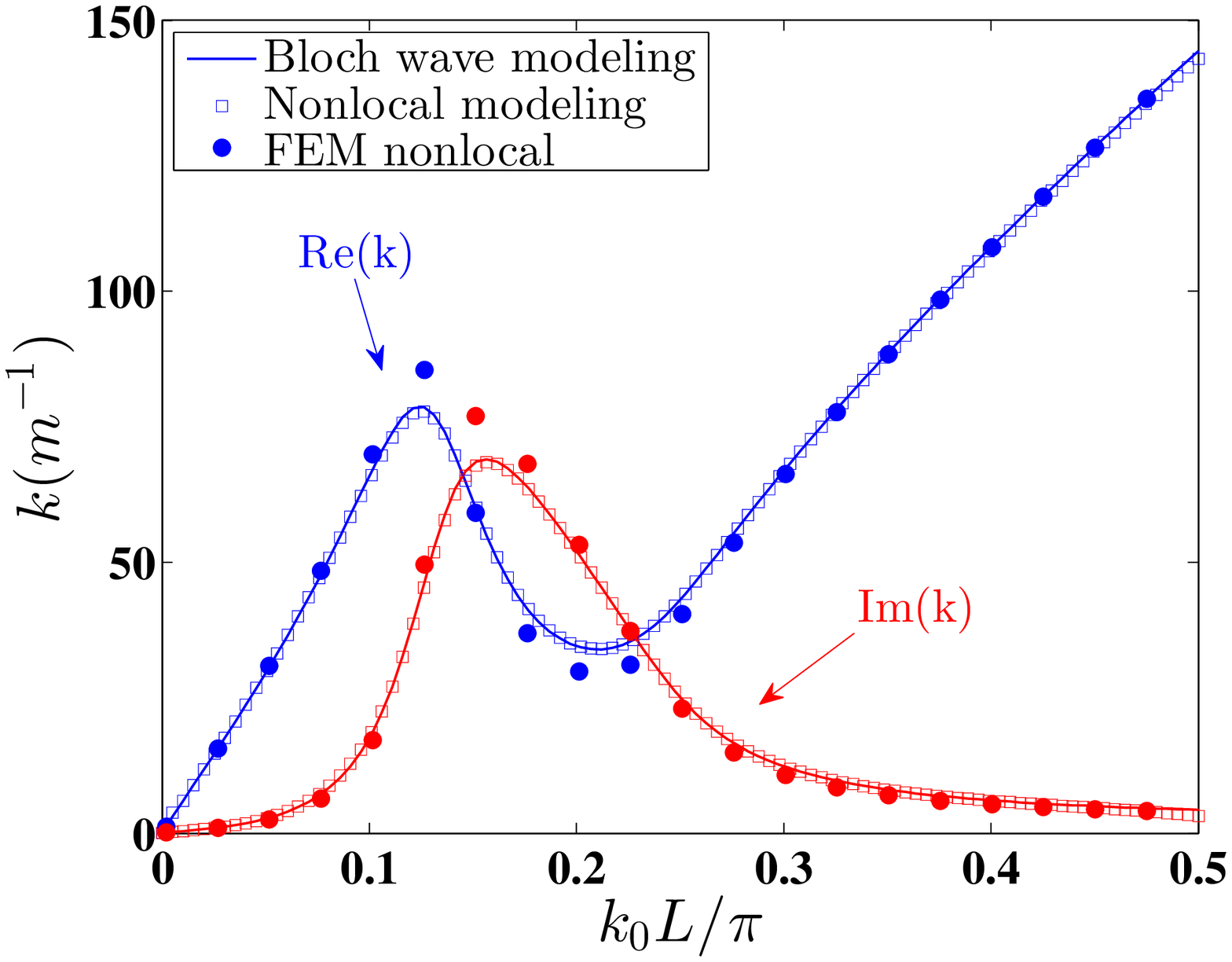}
\includegraphics[width=7.5cm,angle=0]{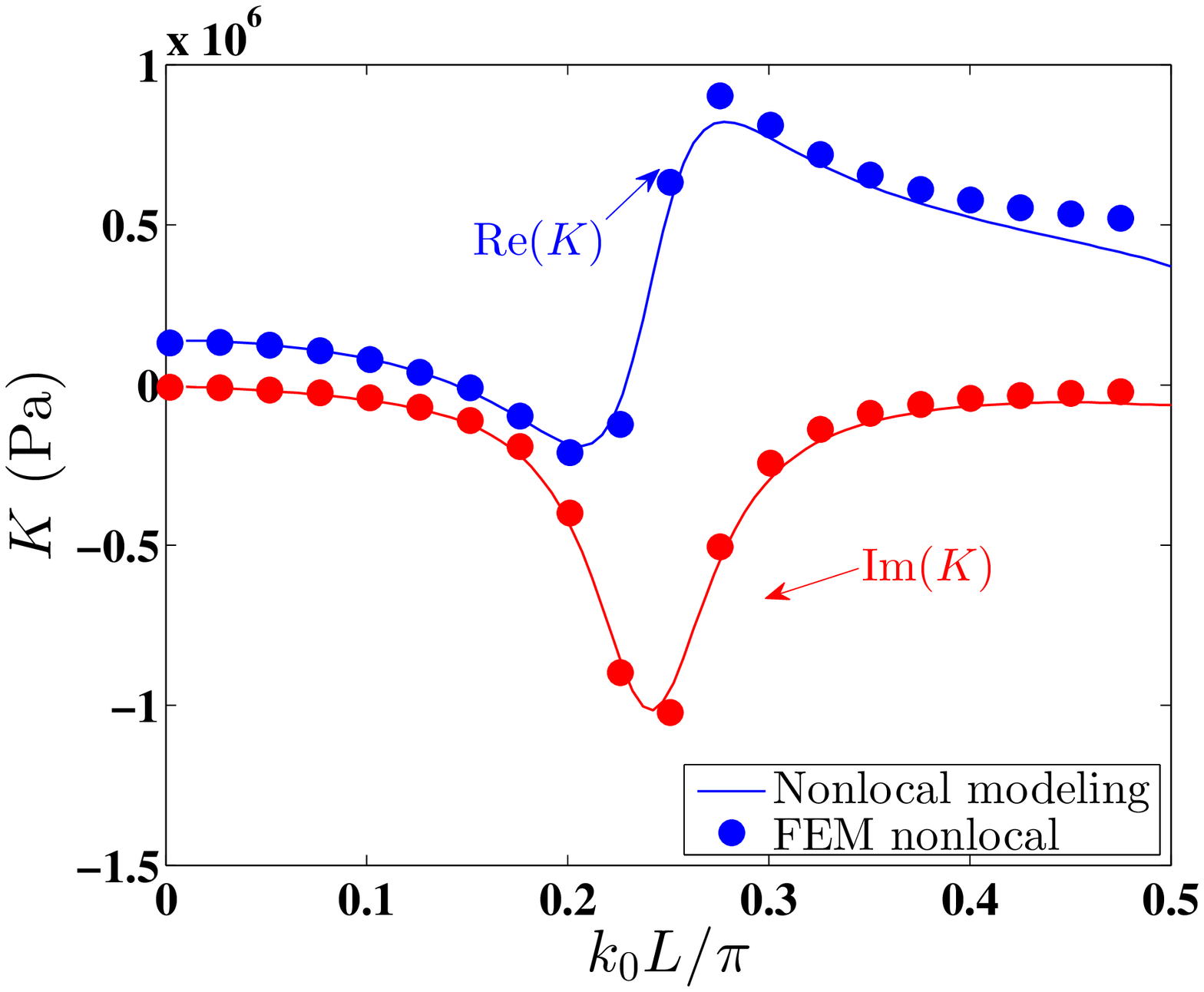}
\end{center}
\caption{Wavenumber (left) and bulk modulus (right) in terms of a dimensionless frequency, for the 2D structure filled with air. For the wavenumber, results by three calculations are compared: Bloch-wave modeling, nonlocal modeling, and nonlocal theory by FEM.}
\label{2Dair}
\end{figure}

We see in Fig.\ref{2Dair}, left, that the real and imaginary parts of $k(\omega)$ computed by nonlocal theory via Newton's method converges exactly to the real and imaginary parts  of $k_{B}$ which has been computed by a simple Bloch-wave modeling without any use of nonlocal theory. The horizontal axis is the dimensionless frequency $k_{0}L/\pi$, where $k_{0}=\omega/c_{0}$.
The results based on the FEM simulations are also in good agreement with those obtained by the Bloch wave modeling and nonlocal modeling. The frequency range has been chosen so that it covers the resonance regime. In the same frequency range, Fig.\ref{2Dair}, right, shows the real and imaginary parts of the effective bulk modulus, computed by nonlocal FEM simulations and nonlocal modeling. Here also, we see excellent agreement between the two calculations. We notice the metamaterial behavior demonstrated in the real part of effective bulk modulus which becomes negative in a frequency range within the resonance regime. It is clear that the results by FEM computations based on the exact microscopic equations, can be considered more precise compared with our two modeling results in which we have applied simplifying approximations. As such, the good agreement between FEM results and others, validate the modeling framework. The discrepancies between the results based on the models and FEM simulations can be due in particular, to the fact that the model describes the admittance of the resonator $Y_{r}$, without considering the length correction of the neck; what might generate errors in the calculation of the wavenumber. 

We observe here the same kind of behavior for the wavenumber and bulk modulus as it has been demonstrated experimentally in \cite{fang} (see Figs. 1 and 2 in that reference) for the case of the 3D material embedded in water. We have observed that removing the thermal effects by decreasing the coefficient of thermal conductivity $\kappa$ to a value close to zero, would have a negligible effect on the wavenumber and the effective bulk modulus. That is the case also for the second viscosity $\zeta$, associated to  losses in the compressional/dilatational motions in the bulk fluid. On the contrary, the material dynamics in terms of the macroscopic wavenumber and bulk modulus is quite sensitive to the values of the shear viscosity $\eta$. In a frequency range, for instance, $k_{0}L/\pi=0.1$ and $0.4$, a maximum and minimum appear for the real part of the wavenumber. By decreasing the value of the shear viscosity, the maximum becomes sharper and finally diverges as the viscosity tends to zero, at the resonance frequency of the ideal fluid $\omega_{H}=c_{0} \sqrt{\sigma/[l (L-2l)(L-\Sigma-2l)]}$, namely $k_{0}L/\pi=0.15$ here; the minimum flattens and  a band gap is created.  As a matter of fact, the important feature here is the resonant behaviour of the structure which induces important values of the velocity in the neck, and thus also important viscous dissipation. Near resonance from below, and at small enough $\eta$, the corresponding neck flow become predominant and the effective wavelength is drastically reduced, leading to a so-called slow speed; but when the shear viscosity increases, the neck flow adjusts to a smaller value, eventually leading to the disappearance of the slow speed.  The viscous losses also smooth out the extrema of the real and imaginary parts of the modulus in Fig. \ref{2Dair} right. Consequently, a wider frequency range of the negative real part of the bulk modulus is obtained by increasing the viscous losses. The thermal boundary layers near the cavity walls, where  the fluid bulk modulus passes from adiabatic to isothermal value,  mainly bring a small correction to the cavity spring constant (the cavity dimension is much greater than the boundary layer thickness). Therefore, their presence do not affect much the effective bulk modulus.

As explained before the dynamics of the material will be very sensitive to the width of the neck, where a considerable part of the viscous losses takes place.  Between the frequencies  $k_{0}L/\pi=0.1$ and $0.4$, the ratio of the viscous boundary layer thickness to the width of the neck, insensibly changes from $0.35$ to $0.39$. We observed that to keep the similar behavior of the wavenumber and modulus, this ratio should remain in the same order, regardless of changing the scale of the material or the saturating fluid. The wavelength in air remains at least about $5$ times larger than the periodicity $L$ where $k_{0}L/\pi=0.5$, and the effective wavelength $\lambda_{eff}$ in the material decreases to $\lambda_{eff}/L\sim 8$  at the resonance frequency $k_{0}L/\pi=0.15$, and to $\lambda_{eff}/L\sim 5$ at $k_{0}L/\pi=0.5$. Although this structure represents a subwavelength material, and can be regarded in the large wavelength limit  $\lambda_{eff}\gg L$,  the local theory based on the two-scale homogenization at order zero does not predict correctly the acoustics of this material. The origin of the failure is the presence of widely different length scales, allowing for resonances. 

Once the simplifying assumptions within our two modeling have been validated by the precise results of the FEM simulations, we can use the same modeling framework to treat the case of 3D material.


\subsection{3D structure filled with air}

Here, the resonators are placed in a periodicity  $L= 1$ cm, composed  of a rectangular cavity of volume $8.5\times5\times5$ mm, a cylindrical neck $l=1$ mm long and $\sigma=1$ mm in diameter, and a main duct portion. The neck opens in the main square air duct with a $\Sigma \times \Sigma=0.2L \times 0.2L$ mm opening. The strategy of calculation to obtain  the effective density, effective bulk modulus and the least attenuated wavenumber through nonlocal modeling and Bloch wave modeling, are the same as for 2D case in sections \ref{nmodeling} and \ref{bmodeling}. We can consider that the $z$-axis is outward from the plane of the Fig. \ref{amplitude} which is regarded as a cross section of the  3D periodic unit. As before, Zwikker and Kosten's plane waves are propagating and attenuating in the different parts of the geometry. The only change which should be applied in the 3D calculations with respect to 2D model,  is related to the Zwikker and Kosten's density and modulus   which have been expressed for slits in Eq. (\ref{slit}). Here, we use the expressions (80) and (81) in \cite{stinson} to obtain the Zwikker and Kosten's density and bulk modulus for tubes of rectangular cross section (main conduit and cavity); for the neck (tube of circular cross section) we use the expression mentioned in  \cite[Appendix]{nemati2014}, and also in \cite{allard,stinson}.  

Fig. \ref{3Dair} left, shows the the real and imaginary parts of the frequency dependent wavenumber $k(\omega)$ associated with the least attenuted mode. The results based on the calculations of nonlocal modeling and Bloch wave modeling appear to be in perfect agreement. In Fig. \ref{3Dair} right, the real and imaginary parts of the frequency dependent bulk modulus $K(\omega)=\chi^{-1}(\omega,k(\omega))$ are presented, according to nonlocal modeling.  

Between the frequencies  $k_{0}L/\pi=0.05$ and $0.3$, the ratio of the viscous boundary layer to the diameter of the neck, changes from $0.15$ to $0.06$. The wavelength in air remains at least about $5$ times larger than the periodicity $L$ where $k_{0}L/\pi=0.5$, and the effective wavelength  in the material decreases to $\lambda_{eff}/L\sim 10$  at the resonance frequency $k_{0}L/\pi=0.07$, and to $\lambda_{eff}/L\sim 5$ at $k_{0}L/\pi=0.5$.

\begin{figure}[h!]
\begin{center}
 \includegraphics[width=6cm]{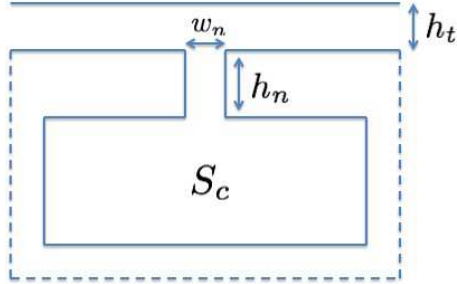}  
\end{center}
\caption{Schematic of the 2D periodic unit of an array of Helmholtz resonator. Shown here are the geometrical parameters, which should be obtained in order to have a 2D equivalent of the 3D structure.}
\label{2D3D}
\end{figure}
\begin{figure}[h!]
\begin{center}
\includegraphics[width=7.7cm,angle=0]{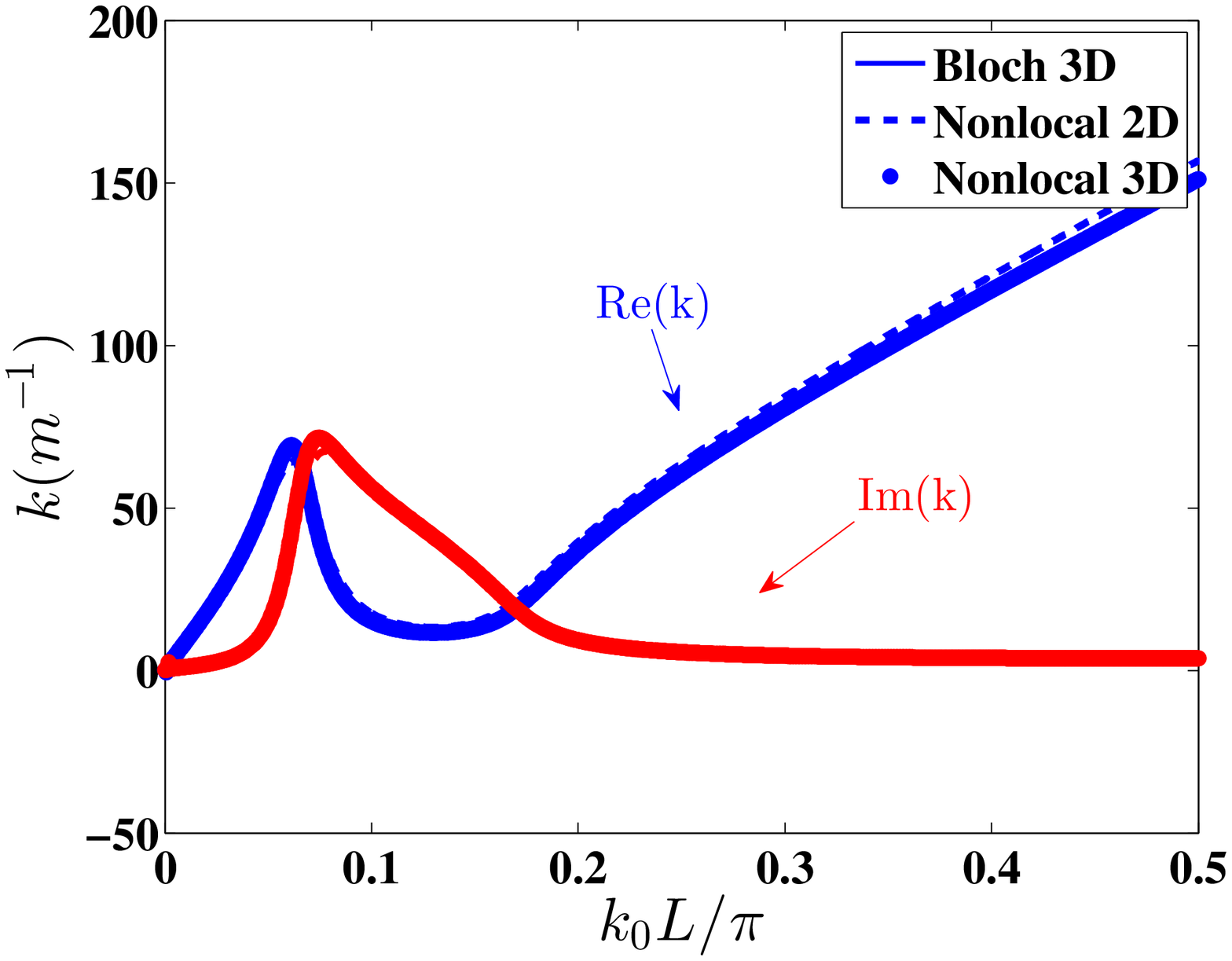}
\includegraphics[width=7.7cm,angle=0]{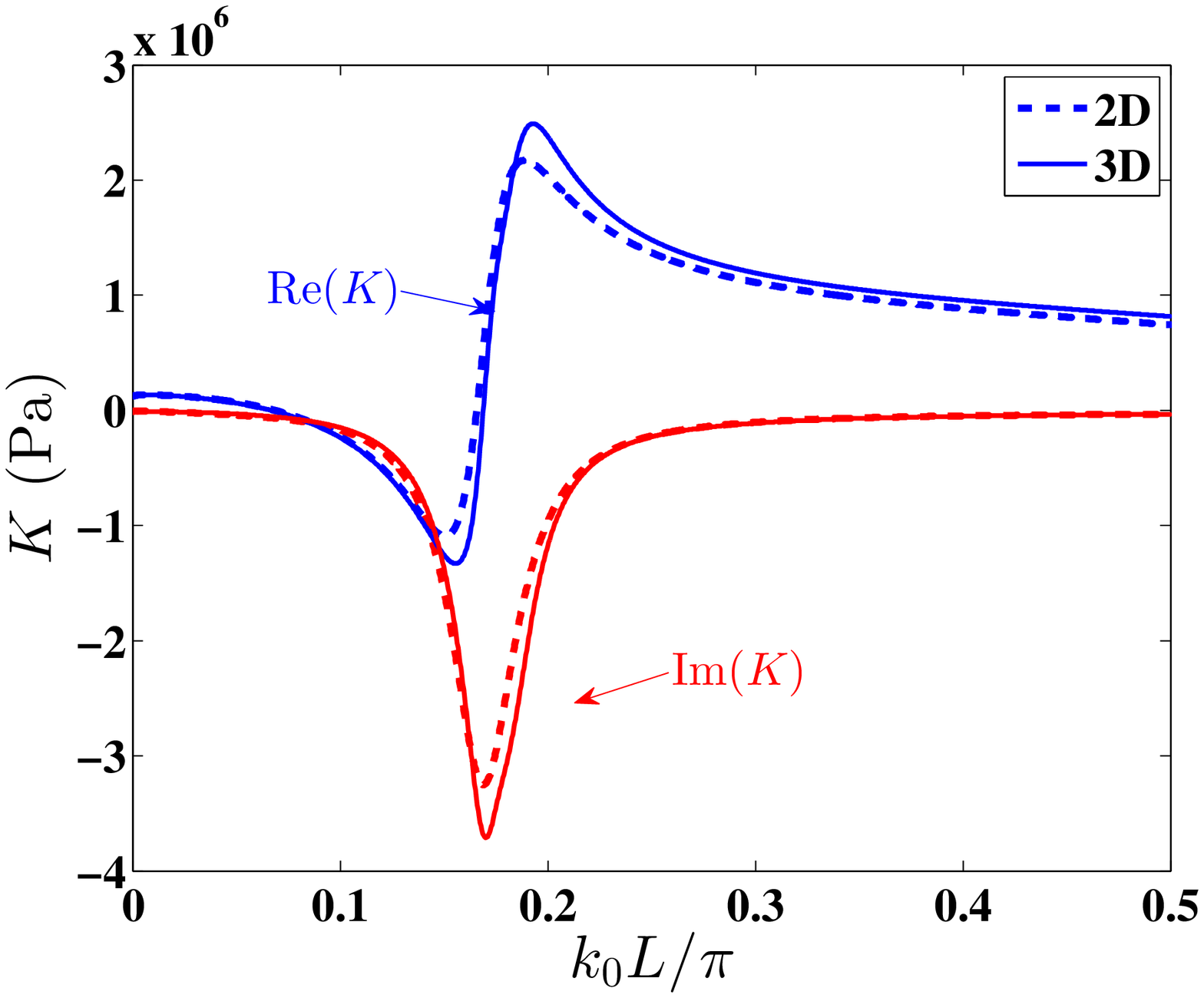}
\end{center}
\caption{Wavenumber (left) and bulk modulus (right) in terms of a dimensionless frequency, for the 3D structure filled with air. For the wavenumber, results by two calculations are compared: Bloch-wave modeling, nonlocal modeling.}
\label{3Dair}
\end{figure}

\textbf{2D equivalent of the 3D structure:} We have performed a simple calculation to obtain the 2D structure made of Helmholtz resonators showing roughly the same macroscopic behavior as a 3D structure, in particular, in terms of the wavenumber of the least attenuated mode and the effective bulk modulus. We will  determine the geometrical parameters of the 2D resonator, illustrated in Fig. \ref{2D3D}, in terms of the parameters of a 3D resonator,  in a way that it exhibits  resonance at the same frequency and shows approximately the same dissipative character as the 3D structure does.    

To have roughly the same amount of both viscous and thermal losses in the main tube in 2D and 3D, it suffices to equate the hydraulic radius (see \cite{allard}). Let $h_{t}^{2D}$ be the width of the tube in 2D, and $h_{t}^{3D}$  
the side of the square tube cross section in 3D. For the hydraulic radius to be the same we take:  $h_{t}^{2D}=h^{3D}_{t}/2$. In the same way, equating the hydraulic radius for the neck in 2D and 3D, gives the neck's width in 2D, $w_{n}^{2D}$ in terms of the diameter of the circular neck: $w_{n}^{2D}=w_{n}^{3D}/2$. The  surface of the cavity in 2D, $S_{c}^{2D}$ is determined in an intutive manner by assuming that the ratio of the cavity volume $V^{3D}_{c}$ to the tube volume $V^{3D}_{t}$ in 3D is equal to the ratio of the cavity surface $S_{c}^{2D}$ to the tube surface $S_{t}^{2D}$ in 2D : $V^{3D}_{c}/V^{3D}_{t}=S_{c}^{2D}/S_{t}^{2D}$. We will have $S_{c}^{2D}=V_{c}\;h_{t}^{2D}/(h_{t}^{3D})^{2}$.
Finally, the equality of the resonance frequency in 2D, $\omega_{H}^{2D} =c_{0} \sqrt{w_{n}^{2D}/(h_{n}^{2D}S^{2D}_{c})}$, and in 3D, 
$\omega_{H}^{3D}  =c_{0} \sqrt{S_{n}^{3D}/(h_{n}^{3D}V^{3D}_{c})}$, results in the expression for the neck's length in 2D: 
$h_{n}^{2D}=(V^{3D}_{c}\;h_{n}^{3D}\;w_{n}^{3D})/(S_{n}^{3D}\;S_{c}^{2D})$, where $h_{n}^{3D}$ is the length of the neck in the 3D structure.

With the  dimensions of our actual 3D structure, we find, for the geometrical parameters of the 2D version,  $ h_{t}^{2D}=1$mm, $w_{n}^{2D}=0.24$mm, $S_{c}^{2D}$, and  $h_{n}^{2D}=8$mm. We have chosen the width $w^{2D}_{c}=8.5$mm and the height $h^{2D}_{c}=6.25$mm, so that the product of them be fixed by the value of $S^{2D}_{c}$. We have also taken the same periodicity $L$ for 2D, as in 3D. The complex wavenumber associated with the least attenuated mode, and complex effective bulk modulus of this 2D equivalent of the 3D material is depicted in Fig. \ref{3Dair}. The complex wavenumber relating to the 2D and 3D geometries present an excellent agreement, and a very good agreement is observed regarding the real and imaginary parts of the effective bulk modulus of these two structures.   

We note that, if the structure with the same geometrical parameters is embedded in water,  there would be less loss as the the viscous boundrary layer thickness is smaller compared with air. To keep the same dynamic behavior with water as with air, it would be necessary to very significantly decrease the width of the neck; at this point it should be born in mind that the complicate effect of nonlinearities would certainly have to be introduced.


Note that the thermal effects in water are not important. The general thermodynamic identity $\gamma-1=\beta_{0}^{2}T_{0}/ \rho_{0}c_{p}$, shows that the deviation of $\gamma\equiv c_{p}/c_{v}$ from unity, is a second order effect on the thermal expansion coefficient $\beta_{0}$ . For a liquid, like water,  $\beta_{0}$ is very small; what implies that $\gamma$ is practically $1$. In this case, isothermal and adiabatic bulk modulus coincide since in general $K_{0(adiab)}=\gamma K_{0(isoth)}$; thermal exchanges have virtually no effects.


\section{Conclusion}


Applying the Maxwellian nonlocal theory of sound propagation in porous media to a material with the microgeometry of the porous matrix  in the form of a two or three dimensional  array of  Helmholtz resonators embedded in air, we have described precisely the metamaterial behavior of the dissipative medium, demonstrated by the negative real part of the effective bulk modulus in the resonance frequency regime. Using the homogenization method corresponding to the recently developed nonlocal theory, we took advantage of a plane wave modelling  to obtain the effective density and bulk modulus, functions of both frequency and wavenumber. In this modeling we made use of Zwikker and Kosten's equations, governing the pressure and velocity fields' dynamics averaged over the cross-section of the different parts of the Helmholtz resonators, in order to coarse-grain them  to the scale of the periodic cell containing one resonator. Once these two effective parameters have been determined, the corresponding least attenuated wavenumber of the medium could be obtained through a dispersion equation established via nonlocal theory. The frequency range has been chosen such that the geometrical-based resonance phenomena could appear. 

On the other hand, a direct analytical modelling has also been performed to obtain the least attenuated Bloch mode propagating in the medium, without using nonlocal theory. We have shown that the values of Bloch modes obtained in the direct way, match exactly those computed by the nonlocal modeling. In addition, the FEM numerical simulations allowing to compute the effective parameters and wavenumbers without any approximation, validate the results of the two modeling calculations and their simplifying assumptions. The nonlocal theory takes fully into account all viscous and thermal dissipation. But we have observed that for this material, thermal effects are negligible, while viscous effects are quite important to describe the material effective dynamics. We have used the same modeling framework for 3D material to compute the effective parameters and the wavenumber of the least attenuated mode, and performed a simple calculation to find a 2D equivalent of the material, showing the same macroscopic dynamics. 

Finally, the resonance induced metamaterial behavior that we have studied here, can be interpretated as a demonstration of the importance of considering the spatial dispersion in the medium.  Higher order modes propagating and attenuating in this material  can also be computed by the nonlocal theory and the subject of future research. Also, the case of the material filled with water will be analysed.








\end{document}